\documentclass[multphys,vecphys]{svmult}

\usepackage{makeidx}         % allows index generation
\usepackage{graphicx}        % standard LaTeX graphics tool
\usepackage{multicol}        % used for the two-column index
\usepackage[bottom]{footmisc}% places footnotes at page bottom
\makeindex             % used for the subject index
\newcommand{\rd}{{\rm d}}
\newcommand{\be}{\begin{equation}}
\newcommand{\ee}{\end{equation}}

\newcommand{\eqn}{\begin{eqnarray}}
\newcommand{\eeqn}{\end{eqnarray}}

\newcommand{\bp}{{\bf p}}
\newcommand{\bv}{{\bf v}}

\newcommand{\Tor}{{\bf T}}
\newcommand{\Fou}{{\cal F}}

\newcommand{\tp}{{\tilde p}}
\newcommand{\tsi}{{\tilde \sigma}}

\newcommand{\tbp}{{\tilde{\bf p}}}

\renewcommand{\a}{\alpha}

\newcommand{\e}{\varepsilon}

\newcommand{\om}{{\omega}}

\newcommand{\bE}{{\bf E}}
\newcommand{\cX}{{\cal X}}
\newcommand{\iint}{\int \!\! \int}
\newcommand{\bR}{{\bf R}}
\newcommand{\bC}{{\bf C}}

\newcommand{\bZ}{{\bf Z}}

\newcommand{\cI}{{\cal I}}
\newcommand{\cH}{{\cal H}}
\newcommand{\cT}{{\cal T}}
\newcommand{\cV}{{\cal V}}
\newcommand{\cO}{{\cal O}}
\newcommand{\cP}{{\cal P}}
\newcommand{\cE}{{\cal E}}

\newcommand{\wt}{\widetilde}
\newcommand{\wh}{\widehat}
\newcommand{\ov}{\overline}

\input epsf

\begin{document}

\title*{Towards the quantum Brownian motion}
\author{L\'aszl\'o Erd\H os\inst{1}\and
 Manfred  Salmhofer\inst{2}\and
 Horng-Tzer Yau\inst{3}}
\institute{Institute of Mathematics, University of Munich,
Theresienstr. 39, D-80333 Munich,
\texttt{lerdos@mathematik.uni-muenchen.de}
\and Theoretical Physics, University of Leipzig, 
Augustusplatz 10, D-04109 Leipzig, 
and Max--Planck Institute for Mathematics, Inselstr.\  22,
D-04103 Leipzig, \texttt{Manfred.Salmhofer@itp.uni-leipzig.de}
\and Department of Mathematics, Stanford University, CA-94305, USA.
 \texttt{yau@math.stanford.edu} }
\maketitle

\abstract{
We consider  random Schr\"odinger equations
on $\bR^d$ or $\bZ^d$ for $d\ge 3$
with uncorrelated, identically distributed random potential.
Denote by $\lambda$ the coupling constant  and $\psi_t$ the solution
with initial data $\psi_0$. Suppose that the space and time variables
scale as $x\sim \lambda^{-2 -\kappa/2}, t \sim \lambda^{-2 -\kappa}$
with $0< \kappa \leq \kappa_0$, where $\kappa_0$ is a sufficiently
small universal constant. 
We prove that
the expectation value of the Wigner distribution of $\psi_t$,
$\bE  W_{\psi_{t}} (x, v)$,  converges weakly to a 
solution of a heat equation
in the space variable $x$ for arbitrary $L^2$  initial data in the weak coupling
limit $\lambda \to 0$.
The diffusion coefficient  is uniquely determined
by the kinetic energy associated to the momentum $v$.}

\section{Introduction}

Brown observed almost two centuries ago
that the motion of a pollen suspended
in water was erratic. This led to the kinetic explanation by
Einstein in 1905 that Brownian motion was created by the
constant ``kicks" on the relatively heavy pollen  by the
light water molecules. Einstein's theory, based upon Newtonian
dynamics of the particles, in fact postulated
the emergence of the Brownian motion from a classical non-dissipative
reversible dynamics. Einstein's theory became universally
accepted after the experimental verification  by Perrin in 1908,
but it was far from being mathematically rigorous.

The key difficulty is similar to the justification of Boltzmann's
molecular chaos assumption (Sto{\ss}zahlansatz) standing behind
Boltzmann's derivation of the Boltzmann equation. The point is that the
dissipative character emerges only in a scaling limit,
as the number of degrees of freedom
goes to infinity.

The first mathematical definition of the Brownian motion was
given in 1923  by Wiener, who  constructed the Brownian motion
as a scaling limit of random walks. This construction
was built upon a stochastic microscopic dynamics
 which by itself are dissipative.

The derivation of the Brownian motion from a Hamiltonian dynamics
was not seriously investigated until the end of the seventies,
when several results came out almost simultaneously.
Kesten and Papanicolaou  \cite{KP} proved that the velocity distribution
of a particle moving in a
random scatterer environment (so-called Lorenz gas with random scatterers)
converges to the Brownian motion in a weak coupling limit for $d\ge 3$.
The same result was obtained 
in $d=2$ dimensions by D\"urr, Goldstein and
Lebowitz \cite{DGL2}.
In this model the bath of light particles is
replaced with  random static impurities.
 In a very recent work \cite{KR}, Komorowski and Ryzhik 
have controlled the same evolution on a longer time scale and proved
the convergence to Brownian motion of the position process as well.

Bunimovich-Sinai \cite{BS} proved the convergence of the
periodic Lorenz gas with a hard core interaction to a Brownian motion. 
In this model the only source of randomness is the distribution
of the initial condition.
Finally, D\"urr-Goldstein-Lebowitz \cite{DGL1} proved that
the velocity process of a
heavy particle in a light ideal gas converges to the
Ornstein-Uhlenbeck process that is a version of the Brownian motion.
This model is the closest to the one in
 Einstein's kinetic argument.

An analogous development happened around the same
time towards the rigorous derivation of the Boltzmann equation.
It was proved by Gallavotti \cite{G}, Spohn \cite{Sp2} and
Boldrighini, Bunimovich and Sinai \cite{BBS}
that the dynamics of the Lorenz gas with random scatterers
converges to the linear  Boltzmann equation at low density
on the kinetic time scale.  Lanford \cite{L} has proved
that a truly many-body classical system, a low density gas
with hard-core interaction, converges to the nonlinear Boltzmann
equation for short macroscopic times.

Brownian motion was discovered and theorized in the context of
classical dynamics.  Since it postulates a microscopic Newtonian
model for atoms and molecules, it is natural to replace
the Newtonian dynamics with the Schr\"odinger dynamics
and investigate if Brownian motion  correctly describes  the motion of
a quantum particle in a random environment as well.
One may of course take first the semiclassical limit, reduce
the problem to the classical dynamics and then consider
the scaling limit. This argument, however, does not apply
to particles (or Lorenz scatterers) of size comparable
with the Planck scale.  It is physically more realistic
and technically considerably more challenging
to investigate the scaling limit of the quantum
dynamics directly {\it 
without any semiclassical limit.} We shall prove
that Brownian motion also  describes the motion of
a quantum particle in this situation.
It is remarkable that the Schr\"odinger evolution, which
is time reversible and describes wave phenomena, converges to a
Brownian motion.

\bigskip

The random Schr\"odinger equation, or the quantum Lorentz
model,  is given  by the evolution equation:
\be\label{sch}
i \partial_t \psi_t(x)=  H \psi_t(x), \qquad
 H=H_\om  = -  \frac 1 2 \Delta_x +  \lambda V_\om(x)
\ee
where $\lambda>0$ is the coupling constant and $V_\om$
is the random potential. 

The first time  scale with a non-trivial limiting
dynamics is the weak coupling limit, $\lambda\to0$,
 where  space and time
are subject to kinetic scaling and the coupling
constant  scales as
\be\label{wc}
t \to t \e^{-1}, \quad x \to x \e^{-1}, \quad \lambda= \sqrt \e \; .
\ee
Under this limit, the appropriately
rescaled  phase space density (Wigner distribution, see (\ref{wig}) later)
 of the solution to
the Schr\"odinger evolution (\ref{sch}) converges weakly to a linear
 Boltzmann equation. 
 This was first established by Spohn (1977) \cite{Sp1}  if the random
potential is a Gaussian random
field and the macroscopic time
is small. This method was extended to  study higher order correlations
by Ho, Landau and Wilkins \cite{HLW}.
 A different method was developed in \cite{EY}
where the short time restriction was removed. This method was also extended
to the phonon case in \cite{E} and to the lattice case in \cite{Ch}.

For longer time scales, one expects a diffusive dynamics since
the long time limit of a Boltzmann equation is a heat equation.
We shall therefore take a time scale  longer
than in the weak coupling limit (\ref{wc}),
 i.e. we set $t\sim \lambda^{-2-\kappa}$,
$\kappa>0$.
Our aim is to prove that the limiting dynamics of the Schr\"odinger evolution
in a random potential under this scaling is governed by a heat equation.
This problem requires to control the Schr\"odinger dynamics
up to a time scale $\lambda^{-2-\kappa}$.
This is a much harder task than first deriving the Boltzmann equation
from Schr\"odinger dynamics on the kinetic scale and then showing
that Boltzmann equation converges to a diffusive equation under
a different limiting procedure. Quantum correlations that are small
on the kinetic scale and are neglected in the first limit,
may contribute on the longer  time scale.

We consider two models in parallel. In the discrete setup
we put the Schr\"odinger equation (\ref{sch})
on $\bZ^d$, i.e. we work with the Anderson model \cite{A}.
Thus the kinetic energy operator on $\ell^2(\bZ^d)$
 is given by
\be\label{Delta}
    (\Delta f)(x): = 2d \; f(x)-\sum_{|e|=1}f(x+e)
\ee
and the random potential is given by
\be\label{ranpot}
        V_\om(x) = \sum_{\gamma\in \bZ^d} V_\gamma(x)\; , \qquad
        V_\gamma(x):= v_\gamma \delta(x-\gamma)
\ee
where $v_\gamma$ are real i.i.d. random variables and $\delta$ is
the lattice delta function, $\delta(0)=1$ and $\delta(y)=0$, $y\neq0$.

In the continuum model we consider the usual Laplacian, 
$-\frac{1}{2}\Delta_x$,
as the kinetic energy operator on $L^2(\bR^d)$. The random potential is
given by
\be\label{ranpotcon}
        V_\omega(x) = \int_{\bR^d} B(x-y)\rd\mu_\omega(y),
\ee
where $\mu_\omega$ is a Poisson process $\{ y_\gamma \; : \;
\gamma=1,2, \ldots \}$ on $\bR^d$ with unit density and i.i.d.
random masses, $v_\gamma$, i.e.
 $\mu_\omega = \sum_{\gamma} v_\gamma\delta(\cdot - y_\gamma)$, and
 $B:\bR^d\to \bR$
is a smooth, radially symmetric function with rapid  decay,
 with $0$ in the support of $\widehat B$.

Since we investigate large distance phenomena, there should be no
physical difference between the continuum and discrete models.
On the technical level, the discrete model is more complicated
due to the non-convexity of the energy surfaces of the
discrete Laplacian in momentum space.  However, 
the continuum model also has an additional technical difficulty:
the large momentum regime needs a separate treatment.

Our proof builds upon 
the method initiated in \cite{EY}. In that paper
the continuum model  with a Gaussian 
random field  was considered. Here we also consider
the discrete model and non-Gaussian randomness, in order to demonstrate
that these restrictions are not essential.
On the Boltzmann scale this extension has also been achieved by
Chen \cite{Ch}.  The other reason for working on the lattice
as well is to make  a connection with the
extended state conjecture in the Anderson model.

We recall that the  Anderson model was invented to describe
the electric conduction properties of disordered metals.
It was postulated by Anderson that for localized initial data
the wave functions for large time are localized  for large coupling constant
$\lambda$ and are extended 
for small coupling constant (away from the band edges
and in dimension $d \ge 3$).
The localization conjecture
was first established rigorously by
Goldsheid, Molchanov and Pastur \cite{GMP} in
one dimension, by Fr\"ohlich-Spencer \cite{FS},
and later by Aizenman-Molchanov \cite{AM} in several
dimensions, and many other works
have since contributed to this field.
The extended state conjecture, however,
has  remained a difficult open problem and only very limited progress
has been made.

Most approaches on extended states focused on the spectral property
of the random Hamiltonian. It was proved by Klein \cite{Kl} that
all eigenfuctions are extended on the Bethe lattice. In Euclidean space,
Schlag, Shubin  and Wolff \cite{SSW}
proved that the eigenfunctions cannot be localized in a region smaller
than $\lambda^{-2+ \delta}$ for some $\delta > 0$ in $d=2$. Chen \cite{Ch}, 
extending the method of \cite{EY} to the lattice case,  proved
that the eigenfunctions cannot be localized
in a region smaller than $\lambda^{-2}$ in any dimension $d\ge 2$
with logarithmic corrections. Lukkarinen and Spohn \cite{LS}
have employed a similar technique for studying energy transport
in a harmonic crystal with weakly perturbed random masses.

A special class of random Schr\"odinger equation was proposed to understand
the dynamics in the extended region. Instead of random potential with
i.i.d. random variables, one considers a
random potential $V_\omega(x)$ with a power law decay, i.e.,
$$
    V_{\omega}(x) =
    h(x)       \omega_x   \; ,  \qquad  h(x)  \sim |x|^{- \eta}
$$
where $\omega_x$ are mean zero i.i.d.  random variables
and $\eta > 0$ is a fixed parameter.

If $\eta \ge 1$
a standard scattering argument yields that
for $\lambda$ small enough $H_\omega$ has absolutely continuous spectrum.
Using cancellation properties of the random potential, Rodnianski and Schlag
\cite{RS} have improved the same result to $\eta > 3/4$
in $d \ge 2$ and  recently,
J. Bourgain \cite{B} has extended it to $\eta > 1/2$.
For $\eta>1/2$ the particle becomes essentially ballistic
at large distances and there are only finitely many  effective collisions.

In summary, in all known results \cite{SSW, RS, B, Ch} for the Anderson
model (or its modification) in Euclidean space
the number of effective collisions are finite.
In the scaling of the current work (\ref{scale}),
the number of effective scatterings 
 goes to infinity in the scaling limit,
 as it should be the case if we aim to obtain a Brownian motion.

As in \cite{Ch}, our  dynamical result also implies that
the eigenfunctions cannot be localized
in a region smaller than $\lambda^{-2-\delta}$ for some $\delta > 0$
and dimension $d\ge 3$ (one can choose $\delta=\kappa/2$
with $\kappa$ from Theorem \ref{main}).
 Though this result is the strongest
in the direction of eigenfunction delocalization, we do not focus on
it here.

Our main result is that  the time reversible
Schr\"odinger evolution with random impurities  
on a time scale $\lambda^{-2-\kappa}$ is described by a dissipative
dynamics. In fact, this work  is the first
rigorous result where a heat equation is established from a time dependent
quantum dynamics without first passing through
a semiclassical limit.

In this contribution we explain the result and the key ideas
in an informal manner. The complete proof is given in \cite{ESY}.

\section{Statement of main result}

We consider the discrete and the continuum models in paralell, therefore
we work either on the $d$-dimensional lattice, 
 $\bZ^d$, or on the continuous space, $\bR^d$.
We always assume $d\ge 3$. Let
\be\label{H}
        H_\omega := -\frac{1}{2}\Delta + \lambda V_\omega
\ee
denote a random Schr\"odinger operator acting on $\cH=l^2(\bZ^d)$, or $\cH=L^2(\bR^d)$.
The kinetic energy operator and the random potential are defined in
(\ref{Delta})--(\ref{ranpotcon}). We assume that $\bE v_\gamma=\bE v_\gamma^3
=0$,  $\bE v_\gamma^2=1$ and $\bE v_\gamma^{2d}<\infty$.

In the discrete case, the Fourier transform is given by
$$
    \wh f(p) \equiv (\Fou f) (p) := \sum_{x\in\bZ^d}
    e^{-2\pi ip\cdot x} f(x) \; , \quad 
%\Big( =: \int_{\bZ^d}   e^{-2\pi ip\cdot x} f(x) \rd x \; \Big),
$$
where $p=(p^{(1)},\dots,p^{(d)})\in\Tor^d:=[-\frac{1}{2},
 \frac{1}{2}]^d$. Sometimes an integral notation will be
used for the normalized summation over any lattice $(\delta \bZ)^d$:
$$
    \int (\cdots ) \rd x: = \delta^d\sum_{x\in (\delta \bZ)^d} (\cdots ) \; .
$$
 The inverse Fourier
transform is given by
$$
    (\Fou^{-1}\wh g)(x)
    = \int_{(\Tor/\delta)^d}   \wh g(p) e^{ 2\pi i p\cdot x}
    \rd p \; .
$$

In the continuous case the Fourier transform and its inverse are given by
$$
     (\Fou f) (p)  :=  \int_{\bR^d}
    e^{-2\pi ip\cdot x} f(x) \rd x\;, \qquad
     (\Fou^{-1}\wh g)(x)
    = \int_{\bR^d}   \wh g(p) e^{ 2\pi i p\cdot x}
    \rd p \;.
$$
We will discuss the two cases in parallel, in particular we will
use the unified integral notations $\int (\cdots) \rd x$ and
 $\int (\cdots) \rd p$.
The letters $x,y,z$ will always be used for position space
coordinates (hence elements of $(\delta\bZ)^d$ or $\bR^d$).
The letters $p, q, r, u, v, w$ denote
for $d$-dimensional momentum
variables (elements of $(\Tor/\delta)^d$ or $\bR^d$).

The Fourier transform of the kinetic energy operator  is given by
$$
    \Big(\Fou \Big[-\frac{1}{2}\Delta\Big] f\Big)(p) =   e(p) \wh f(p) \; .
$$
The dispersion law, $e(p)$, is given by
$$
    e(p) := \sum_{i=1}^d (1-\cos( 2\pi p^{(i)})), \qquad\mbox{and}\qquad   e(p):= \frac{1}{2}p^2 
$$
in the discrete and in the continuous case, respectively.

For $h:\Tor^d\to\bC$ 
and an energy value $e\in [0,2d]$ we  introduce the notation
\be\label{coar}
   [ h ](e):=
   \int h(v) \delta(e-e(v))\rd v : = \int_{\Sigma_e} h(q)
   \; \frac{\rd \nu(q)}{|\nabla   e(q)|}
\ee
where $\rd \nu(q)=\rd\nu_e(q)$ 
is the restriction of the $d$-dimensional Lebesgue measure
to the level surface $\Sigma_e:=\{ q\; : \; e(q)=e\}\subset \Tor^d$. 
By the co-area formula it holds that
\be
    \int_0^{2d} [ h ](e) \rd e   = \int h(v) \rd v \; .
\label{coa}
\ee
We define the projection onto the energy space of the free Laplacian by 
\be\label{F}
\langle \, h(v) \, \rangle_e : = \frac{[h](e)}{ \Phi(e)}\; ,
\quad\mbox{ where }\quad
\Phi(e): = [1](e)=\int \delta(e-e(u))\rd u \;.
\ee
In the continuous case we define analogous formulas 
for any function $h:\bR^d\to \bC$ and energy value $e\ge 0$.
%we set
%$$
%    [h](e):= \sqrt{2e} \int_{S^2} h(\sigma\sqrt{2e})d\sigma 
%    \; , \qquad \Phi(e)= 4\pi\sqrt{2e}\; , 
%$$
%where $d\sigma$ is the Lebesgue measure on the unit sphere $S^2$.
%The energy space projection $\langle \, h(v) \, \rangle_e$ is defined
%by (\ref{F}).

Define the {\it Wigner transform} of a function $\psi\in
L^2(\bZ^d)$ or $\psi\in L^2(\bR^d)$ via its Fourier transform by
$$
        W_\psi(x,v):
        = \int e^{2\pi i  w \cdot x}\overline{\widehat
\psi\Big(v-\frac{w}{2}\Big)}
        \widehat\psi \Big(v+\frac{w}{2}\Big) \rd w \;\; .
$$
In the lattice case the integration domain is the double torus
$(2\Tor)^d$ and $x$ runs over the refined lattice, $x\in (\bZ/2)^d $.
For $\e>0$ define the rescaled Wigner distribution as
\be
        W^\e_\psi (X, V) : = \e^{-d}W_\psi\Big( {X\over \e}, V\Big).
\label{wig}
\ee
(with $X\in (\e\bZ/2)^d$ in the lattice case).

\bigskip

The weak coupling limit is defined by the following scaling:
\be \label{wcl}
\cT:=\e t,\quad
\cX:= \e x, \quad \e= \lambda^2\; .
\ee
In the limit $\e\to 0$
the Wigner distribution $W^\e_{\psi_{\e^{-1} \cT}} (\cX, \cV)$
converges weakly to the Boltzmann equation (\cite{EY}, \cite{Ch})
\be
 \Big( \partial_\cT  +
 \frac{1}{2\pi}\nabla e(V)\cdot \nabla_\cX \Big)F_\cT(\cX,V)=
        \int dU  \sigma(U, V) \Big[ F_\cT(\cX,U)
         -  F_\cT(\cX,V) \Big]
\label{B}
\ee
where $  \frac{1}{2\pi}\nabla e(V)$ is the velocity.
The collision kernel is given by
$$
   \sigma(U, V): = 2\pi \delta(e(U)-e(V))    \qquad \mbox{discrete case}
$$
$$
    \sigma(U, V): = 
  2\pi |\wh B(U-V)|^2\delta(e(U)-e(V))    \qquad \mbox{continuous case} \; .
$$
Note that the Boltzmann equation can be viewed as
 the generator of a Markovian semigroup on phase space.
In particular, the validity of the Boltzmann equation
shows that all correlation effects become negligible
in this scaling limit.

\bigskip

Now we consider the long time scaling, i.e. with some
 $\kappa > 0$,
\be\label{scale}
x= \lambda^{-\kappa/2-2}X = \e^{-1} X, \quad t =
\lambda^{-\kappa-2} T = \e^{-1} \lambda^{-\kappa/2} T,
\quad \e = \lambda^{\kappa/2+2}
\ee

\begin{theorem}\label{main} [Quantum Diffusion on Lattice]
Let $d=3$ and $\psi_0 \in \ell^2(\bZ^d)$ be an initial wave function 
with $ \wh\psi_0 \in C^1(\Tor^d)$.
Let $\psi(t)=\psi_{t,\om}^\lambda$ solve the Schr\"odinger equation
(\ref{sch}). Let $\wt \cO(x, v)$ be a function on $\bR^d\times \Tor^d$
whose Fourier transform in the first variable, denoted by
$\cO(\xi, v)$, is a $C^1$ function on $\bR^d\times \Tor^d$ and
\be \label{obound}
\int_{\bR^d} \rd \xi \int \rd v |\cO(\xi, v)||\xi| \le C\; .
\ee
Fix $e\in [0,2d]$. Let $f$ be the solution to the heat equation
\be
\partial_T f(T, X, e) = \nabla_X\cdot D(e)
\nabla_X f(T, X, e) \label{eq:heat}
\ee
with the initial condition
$$
     f(0, X, e): = \delta(X) \Big[ |\wh \psi_0(v)|^2 \Big](e)
$$
and the diffusion matrix
$D$
    \be D_{ij}(e):=  \frac{
    \big\langle \,  \sin (2\pi v^{(i)})\cdot \sin (2\pi v^{(j)})
    \, \big\rangle_e}{2\pi\; \Phi(e)}
\; \qquad  i,j=1,2,3\; .\label{diffconst}
\ee
Then for $\kappa<1/2000$ and $\e$ and $\lambda$ related by (\ref{scale}),
the Wigner distribution  satisfies
\be
\lim_{\e \to 0} \int_{(\e\bZ)^d} \rd X\int_{\bR^d}
\rd v \tilde \cO(X, v)  \bE
 W^\e_{\psi(\lambda^{-\kappa-2} T)} (X, v)\qquad\qquad\qquad\qquad
\label{fint}
\ee
$$
\qquad\qquad\qquad = \int_{\bR^d} \rd X\int_{\bR^d} \rd v \; \tilde \cO(X, v) f(T, X, e(v)) \; .
$$
\end{theorem}
By the symmetry of the measure $\langle \cdot\rangle_e$
under each sign flip $v_j\to -v_j$ 
we see that $D(e)$ is a constant
times  the identity matrix:
$$
    D_{ij} (e) = D_e \; \delta_{ij}, \qquad D_e:=
\frac{ \big\langle  \sin^2 (2\pi v^{(1)})
    \, \big\rangle_e}{2\pi\; \Phi(e)}  \; ,
$$
in particular we see that the diffusion is nondegenerate.

The diffusion matrix can also be obtained from the long time
limit of the Boltzmann equation (\ref{B}). For any fixed energy $e$,
let
\be
   L_e f(v): = \int \rd u \; \sigma(u, v) [ f(u)-f(v)], \qquad e(v)=e\;,
\label{Lgen}
\ee
be the generator of the momentum jump process on $\Sigma_e$
with the uniform stationary measure $\langle \cdot \rangle_e$.
The diffusion matrix in general is  given by the
velocity autocorrelation function
\be
  D_{ij}(e)=\int_0^\infty \rd t  \; 
   \big\langle \sin (2\pi v^{(i)}(t))\cdot \sin(2\pi v^{(j)}(0))\big\rangle_e \; ,
\label{Dij}
\ee
where $v(t)$ is the process generated by $L_e$.
Since the  collision kernel $\sigma(U,V)$ is   uniform,
the correlation between $v(t)$ and $v(0)$  vanishes after
the first jump and we obtain  (\ref{diffconst}), using
$$
    \int\rd u\; \sigma(u, v)  = 2\pi \Phi(e)\; , \qquad e(v)=e\; .
$$

The result in the continuum case is analogous.
The diffusion matrix is again
a constant times the identity matrix, $D_{ij}(e)= D_e \delta_{ij}$,
and $D_e$ is again given by the velocity autocorrelation function
\be
     D_e : 
   =  \frac{1}{3(2\pi)^2}
     \int_0^\infty \rd t \;\langle v(t)\cdot v(0) \rangle_e 
\label{De}
\ee
using the spatial isotropy. In this case $D_e$ cannot
be computed as a simple integral since the outgoing
velocity $u$ in the transition kernel
$\sigma(u,v)$ of the momentum process 
depends on the direction of $v$.

\begin{theorem}\label{mainc} [Quantum Diffusion on $\bR^d$]
Let $d=3$ and $\psi_0 \in L^2(\bR^d)$ be an initial wave function 
with
$|\wh\psi_0(v)|^2 |v|^{N}\in L^2$ for a sufficiently  large $N$.

Let $\psi(t)=\psi_{t,\om}^\lambda$ solve the Schr\"odinger equation
(\ref{sch}). Let $\wt \cO(x, v)$ be a function
whose Fourier transform in $x$, denoted by
$\cO(\xi, v)$, is a $C^1$ function on $\bR^d\times \bR^d$ and
\be \label{oboundc}
\iint \rd \xi \rd v |\cO(\xi, v)||\xi| \le C\; .
\ee
Let $e>0$ and
let $f$ be the solution to the heat equation
\be
\partial_T f(T, X, e) = D_e\; \Delta_Xf(T, X, e) \label{eq:heatc}
\ee
with diffusion constant $D_e$ given in (\ref{De}) and 
with the initial condition
$$
     f(0, X, e): = \delta(X) \Big[ |\wh \psi_0(v)|^2 \Big](e) \; .
$$
Then for $\kappa<1/500$ and $\e$ and $\lambda$ related by (\ref{scale}),
the Wigner distribution  satisfies
\be
\lim_{\e \to 0} \iint_{\bR^d\times\bR^d} \rd X  \rd v \wt \cO(X, v)  \bE
 W^\e_{\psi(\lambda^{-\kappa-2} T)} (X, v)\qquad\qquad\qquad\qquad
\label{fintc}
\ee
$$
\qquad\qquad\qquad
= \iint_{\bR^d\times\bR^d} \rd X  \rd v \; \wt \cO(X, v) f(T, X, e(v)) \; .
$$
\end{theorem}

The main tool of our proof is to use the
Duhamel expansion to decompose the wave function into elementary
wavefuctions characterized by their collision histories
with the random obstacles. Assume for the moment that the
randomness is Gaussian and high order expectations can
be computed by Wick pairing. The higher order cumulants
arising from  a non-Gaussian randomness  
turn out to be negligible by a separate argument.
Therefore, when computing
the expectation of a product involving $\psi$ and $\bar\psi$
(e.g. $\bE\; W_\psi$), we pair the obstacles in
the collision histories of $\psi$ and $\bar\psi$
and we thus generate Feynman graphs.

If we take only the Laplacian as the free part in the expansion,
even the amplitudes of individual graphs diverge in the limit we consider.
However, this can be remedied by a
simple resummation  of all two-legged insertions
caused by the lowest order self-energy contribution
The resummation is performed by choosing an appropriate
reference Hamiltonian $H_0$ for the expansion. 
After this rearrangement, all graphs have a finite amplitude 
in our scaling limit, and the so-called
 ladder graphs give the leading  contribution.

Each non-ladder graph has a vanishing amplitude as $\lambda\to0$
due to oscillatory integrals, in contrast to the ladder
graphs where no oscillation is present. However, the number
of non-ladder graphs
 grows as $k!$, where $k \sim \lambda^2t \sim \lambda^{-\kappa}$
is the typical number of collisions. To beat this  factorial growth, 
we need to give a very sharp bound on the individual graphs.

We give a classification of arbitrary large graphs,
based on counting the number of
vertices carrying oscillatory effects. The number of
these vertices is called  the {\it degree } of the graph.
For the ladder  graphs, the degree is zero.
For general graphs, the degree is roughly
the number of vertices after removing all ladder and anti-ladder
subgraphs. We thus obtain an extra $\lambda^c$ power
(for some $c>0$) {\em
per non-(anti)ladder vertex}. This strong improvement is sufficient
to beat the growth of the combinatorics in the time scale we consider.
To our knowledge, nothing like this has been done in a
graphical expansion before.

For a comparison, the  unperturbed Green functions in the perturbation expansion
for the many-fermion systems for small temperature
 and for the random Schr\" odinger equation for large time are
given  by
$$
\frac 1 {ip_0+p^2-\mu },  \qquad
\frac 1 {p^2-\alpha + i \eta}\; .
$$
In the many-fermion case, 
$p_0 \in M_F = \{\frac{\pi}{\beta} (2 n + 1): n \in \bZ\}$
where $\beta \sim T^{-1}$ is the inverse temperature. 
In the random Schr\" odinger case, $\eta \sim t^{-1}$. 
Their $L^2$ properties  are different:
$$
\frac{1}{\beta} \sum_{p_0 \in M_F}
\int d p  \big |ip_0+p^2-\mu \big |^{-2}  \sim |\log \beta|, \quad
\int d p \big |p^2-\alpha + i \eta \big |^{-2}  \sim \eta^{-1}
$$
Notice the divergence is more severe for the random Schr\"odinger equation case.
%In the many-fermion case, 
%there is one $p_0$--summation per line of the graph; in the random Schr\" odinger case
%there are just two overall $\alpha$--integrals for graphs with arbitrarily many 
%lines.

Finally we note that the threshold $\kappa <1/2000$ in our theorem
can be significantly improved with more detailed arguments.
However,  one cannot go beyond
$\kappa=2$ with only improvements on estimates of the
individual graphs.  The Duhamel formula must be expanded at
least up to $k=\lambda^2t = \lambda^{-\kappa}$,  which  is the typical number
of collisions up to time $t$. Even if one
proves for most
graphs the best possible estimate, $\lambda^{2k}$,
it cannot beat the $k!$ combinatorics when
$k\gg \lambda^{-2}$, i.e., $\lambda^{2k} k! \gg 1 $ for $k\gg \lambda^{-2}$.
A different resummation procedure is needed beyond this threshold
to exploit cancellations among these graphs.

\section{Sketch of the proof}

We present the main ideas of the proof for the lattice case
and comment on the modifications for the continuous case.

\subsection{Renormalization}
Before expanding the solution of the Schr\"odinger equation (\ref{sch})
via the Duhamel formula, we perform a renormalization of  
the "one-particle propagator"
by splitting the Hamiltonian as $H=H_0 + \tilde V$, 
with $H_0$ already containing the part of the self--energy
produced by immediate recollisions with the same obstacle.
This effectively resums all such immediate recollisions.

Let $\theta(p) := \Theta(e(p))$, where
$\Theta (\alpha): = \lim_{\e \to 0+ } \Theta_\e (\alpha)$ and
\be\label{theta}
    \Theta_\e (\alpha): =  \int \frac  {1} { \alpha- e(q) + i \e} \rd q \; . 
\ee
We have 
\begin{equation}
{\rm Im}\; \Theta(\alpha ) =-\pi\Phi(\alpha)
\label{Ith}
\end{equation}
with $\Phi$ defined in (\ref{F}).

We rewrite the Hamiltonian as
$H= H_0 + \wt V$
with
\be\label{renH}
  H_0:=\omega(p):= e(p) +\lambda^2 \theta(p), \qquad \wt V := \lambda
    V -\lambda^2 \theta(p)\; .
\ee
Our renormalization includes only the lowest order 
self--energy. This suffices on the time scales we consider.

\subsection{The Expansion and the Stopping Rules}
Iterating the Duhamel formula
\begin{equation}\label{eq:standuha}
e^{-itH} = e^{-itH_0} - i \int_0^t ds\; e^{-i(t-s)H} \wt V e^{-isH_0}
\end{equation}
gives for any fixed integer $N\ge 1$
\begin{equation}\label{duh}
        \psi_t : = e^{-itH}\psi_0 = \sum_{n=0}^{N-1} \psi_n (t)
     + \Psi_{N}(t) \; ,
\end{equation}
with
\be
        \psi_n(t) : = (-i)^n\int_0^t [\rd s_j]_1^{n+1} \; \;
    e^{-is_{n+1}H_0}\wt V e^{-is_nH_0}\wt V\ldots
       \wt V e^{-is_1 H_0}\psi_0
\label{eq:psin}
\ee
being the fully expanded terms
and
\be
        \Psi_{N} (t): = (-i) \int_0^t \rd s \, e^{-i(t-s)H}
   \wt V \psi_{N-1}(s)
\label{eq:PsiN}
\ee
is the  non-fully expanded or error term. We used the shorthand notation
$$
    \int_0^t [\rd s_j]_1^n : = \int_0^t\ldots \int_0^t
 \Big(\prod_{j=1}^n \rd s_j\Big)
        \delta\Big( t- \sum_{j=1}^n s_j\Big) \; .
$$
Since each potential $\wt V$ in (\ref{eq:psin}), (\ref{eq:PsiN})
is a summation itself, $\wt V=
-\lambda^2\theta(p)+ \lambda \sum_\gamma V_\gamma$,
 both of these terms in (\ref{eq:psin}) and (\ref{eq:PsiN})
are actually big summations over so-called elementary wavefunctions,
which are characterized by their collision history, i.e. by a sequence
of obstacles labelled by $\gamma \in \bZ^d$ and a label $\vartheta$
corresponding to an insertion of $- \lambda^2 \theta(p)$.

Because this expansion is generated by iteration of (\ref{eq:standuha}),
the sequences defining collision histories
can be obtained recursively. This allows us to refine the 
Duhamel expansion by using stopping rules
that depend on the type of collision history. 
We call a sequence {\em nonrepetitive} if the only repetitions in potential 
labels $\gamma$ occur in gates (immediate recollisions).
The iteration  of (\ref{eq:standuha})  is stopped when 
adding a new entry to the sequence makes it violate this condition. 
This can happen because of a recollision, a nested recollision, or
a triple collision.  The precise definition of these recollision types
is given in \cite{ESY}.
The iteration is also stopped when the last entry in the sequence causes the 
total number of gates and $\vartheta$'s  to reach $2$. 
If the sequence stays nonrepetitive and the total number of gates and 
$\vartheta$'s stays below $2$, the iteration is stopped when
the number of non--gate potential labels reaches 
\be
     K = \lambda^{-\delta}(\lambda^2 t)\;.
\label{def:K}
\ee
Note that $K$
is much bigger than the expected typical number of collisions, $\lambda^2t$.

We denote  the sum of the truncated elementary non-repetitive 
wave functions with at most one
$\lambda^2$ power from the non-skeleton indices or $\vartheta$'s
and with $K$ skeleton indices by
$\psi_{*s, K}^{(\leq 1),nr}$. The superscript $(\leq 1)$ refers to
the number of gates and $\vartheta$'s, each of which gives
 a factor $\lambda^2$. 
By this splitting, we arrive at the following modified 
Duhamel formula, in which all non--error terms are nonrepetitive.

\begin{proposition}\label{prop:duh}[Duhamel formula] 
For any $K\ge 1$ we have
\be
\psi_t=e^{-itH}\psi_0 = \sum_{k=0}^{K-1} \psi_{t,k}^{(\leq 1),nr}\qquad 
\qquad \qquad\qquad \qquad \qquad \qquad \qquad \qquad 
\label{eq:duha}
\ee
$$
 + \int_0^t\rd s \; e^{-i(t-s)H}\Bigg\{
\psi_{*s, K}^{(\leq 1),nr}+\sum_{k=0}^{K}
\Big( \psi_{*s,k}^{(2),last} + \psi_{*s,k}^{(\leq 1),rec}+
\psi_{*s,k}^{(\leq 1),nest}+\psi_{*s,k}^{(\leq 1),tri}\Big)
\Bigg\}
$$
\end{proposition}
The terms under the integral correspond to the various stopping criteria indicated
above. For the precise definition of the corresponding wave functions, see \cite{ESY}.

The main contribution comes from the non-repetitive sequences with $k<K$, i.e.
from the first term in (\ref{eq:duha}).
The estimate of the terms in the second line (\ref{eq:duha}) first uses
the unitarity of the full evolution
\be
 \Big\|  \int_0^t\rd s \; e^{-i(t-s)H} \psi_s^\#\Big\| \leq t \cdot 
 \sup_{s\leq t} \| \psi_s^\#\| \; .
\label{crude}
\ee
For $\#=rec, nest, tri$ we will use the fact the Feynman graphs arising
in the expectation $\bE  \| \psi_s^\#\|^2$
contain an additional
oscillatory factor, which renders them smaller than the corresponding
non-repetitive term. It turns out that the oscillation effect
from one single recollision, nest or triple collision is already sufficient
to overcome the additional factor $t$ arising from
the crude bound (\ref{crude}). This fact relies on estimates on
singular integrals concentrating on the energy level sets $\Sigma_e$.
It is a well-known fact from harmonic analysis, that
such singular integrals can more effectively be estimated
for convex level sets. This is why the non-convexity of the energy shells
is a major technical complication for the discrete model in comparison with the continuous
case, where the level sets are spheres.

Non-skeleton labels also give
rise to a smallness effect due to a cancellation between gates and
$\vartheta$'s, however, one such cancellation would not be sufficient
to beat the $t$--factor. This is why at least two such cancellations
are necessary in $ \psi_{*s,k}^{(2),last}$.
Finally, the term $\psi_{*s, K}^{(\leq 1),nr}$ is small because
it has unusually many collisions, thanks to the additional factor $\lambda^{-\delta}$
in the definition of $K$.

In this exposition we focus only on the non-repetitive terms,
$\psi_{t,k}^{(\leq 1),nr}$,
because estimating them involves the main new ideas. 
The error terms are estimated by laborious technical modifications
of these ideas.

\subsection{The $L^2$ norm of the non-repetitive wavefunction}

We first estimate the $L^2$ norm of the fully expanded wave function
with no gates or $\vartheta$, $\psi_{t,k}^{(0),nr}$.
This is the core of our analysis.

\subsubsection{Feynman Graphs}

The wavefunction
$$
    \psi_{t,k}^{(0),nr}=  \sum_{\gamma}
    \int_0^t [\rd s_j]_1^{k+1} \; \; e^{-is_{k+1}H_0}V_{\gamma_k}
    e^{-is_kH_0} V_{\gamma_{k-1}} \ldots
    e^{-is_2H_0} V_{\gamma_1} e^{-is_1H_0} \psi_0
$$
where the summation is over all sequences for which 
the potential labels $\gamma_i$ are all different.
Therefore
every term in
$$
    \bE \|\psi_{t,k}^{(0),nr}\|^2 = \sum_{\gamma, \gamma'}
    \bE\; \int \ov{\psi_{t,\gamma}}\psi_{t,\gamma'} 
$$
has $2k$ potential terms, and their expectation,
$$
    \bE \; \ov{ V_{\gamma_1} V_{\gamma_2} \ldots  V_{\gamma_k}}
    V_{\gamma_1'} V_{\gamma_2'} \ldots  V_{\gamma_k'}\; ,
$$
is zero, using  $\bE V_\gamma =0$, unless the potentials are paired.
Since there is no repetition within $\gamma$ and $\gamma'$,
all these pairings occur between  $\gamma$ and $\gamma'$,
therefore every pairing corresponds to a permutation on
$\{ 1,2, \ldots , k\}$. The set of such permutations is denoted by
 $\cP_k$ and they can be considered as a map between the indices of the 
 $\gamma$ and $\gamma'$ labels.

We recall the following identity from Lemma 3.1 of \cite{EY}
$$
    \int_0^t [\rd s_j]_1^{k+1} \prod_{j=1}^{k+1} e^{-is_j\om(p_j)}
    = \frac{ie^{\eta t}}{2\pi}\int_{\bR} \rd \alpha \; e^{-i\alpha t}  \prod_{j=1}^{k+1}
    \frac{1}{\alpha - \om(p_j)+i\eta}
$$
for any $\eta>0$. We will choose $\eta:=t^{-1}$.
Therefore, we have
\be
    \bE \|\psi_{t,k}^{(0),nr}\|^2 = \frac{\lambda^{2k}e^{2t\eta}}{(2\pi)^2}
    \sum_{\sigma\in \cP_k}
    \sum_{\gamma_1, \ldots ,\gamma_k\atop \gamma_i\neq\gamma_j}
    \int\rd\bp \rd \tbp \;
    \delta(p_{k+1}-\tp_{k+1})\qquad\qquad\qquad
\label{eq:psiM}
\ee
$$
\qquad\qquad\qquad\times  \bE \prod_{j=1}^k  \overline{\wh V_{\gamma_j}(p_{j+1}-p_{j})}
     \wh V_{\gamma_j}( \tp_{\sigma(j)+1}-\tp_{\sigma(j)})
     M(k, \bp,\tbp,\eta)\overline{\wh\psi_0(p_1)}
    \wh\psi_0(\tp_1)
$$
with $\bp = (p_1, p_2, \ldots, p_{k+1})$,
$
    \int \rd\bp: =
    \int_{(\Tor^d)^{k+1}} \rd p_1 \rd p_2\ldots \rd p_{k+1},
$
similarly for $ \tbp$ and $\rd\tbp$,
and
\be
    M_\eta(k, \bp,\tbp):=\int\!\!\int_{\bR}\rd\alpha\rd\beta
    \; e^{i(\alpha-\beta)t}
    \Bigg(\prod_{j=1}^{k+1} \frac{1}{\alpha - \ov\om(p_j)-i\eta}\;
    \frac{1}{\beta -\om(\tp_j)+i\eta}\Bigg)\; .
\label{def:Mi}
\ee
We compute the expectation:
\be
    \bE \prod_{j=1}^k  \overline{\wh V_{\gamma_j}(p_{j+1}-p_{j})}
     \wh V_{\gamma_j}( \tp_{\sigma(j)+1}-\tp_{\sigma(j)})=
    \sum_{\gamma_1, \ldots, \gamma_k\atop \gamma_i\neq\gamma_j}
   \prod_{j=1}^k e^{i\gamma_j(p_{j+1}-p_{j}-
   (\tp_{\sigma(j)+1}-\tp_{\sigma(j)}))}\; .
\label{eq:treegraph}\ee
In the continuous model this formula also contains a product
of $\wh B$-terms, where $B$ was the single site potential function
in (\ref{ranpotcon}). These factors are included into the
definition of $M_\eta$. Most importantly, they  provide
the necessary decay in the momentum variables in 
the case of non-compact momentum space. A similar
idea was used in \cite{EY}.

Due to the restriction $\gamma_i\neq\gamma_j$, (\ref{eq:treegraph}) is not a simple
product of delta functions in the momenta.
We have to  use a connected graph expansion
that is well known in the polymer expansions of field theory (see, e.g. \cite{S}).
We do not give the details here, we only note that the result is a weighted 
sum over partitions of the index set
 $\{1, \ldots , k\}$. Each term in the sum is a product
 of delta functions labelled by the lumps of the partition and
each delta funtion imposes the Kirchoff Law
for the incoming and outgoing momenta of the lump and its $\sigma$-image.
The trivial partition, where each lump has a single element, carries the
main contribution.
Estimating the terms with nontrivial partitions 
 can be reduced to estimates for the trivial partition
\cite{ESY}. We therefore discuss only the contribution from the trivial partition
to $\bE \|\psi_{t,k}^{(0),nr}\|^2$, given by
$\sum_{\sigma\in\cP_k} V_\eta(k,\sigma)$, where
$$
      V_\eta(k,\sigma): =  \frac{\lambda^{2k}
 e^{2t\eta}}{(2\pi)^2}\int \rd\bp  \rd\tbp
      M_\eta(k,\bp,\tbp) \delta( \tp_{k+1}-p_{k+1}) 
     \overline{\wh\psi_0(p_1)}\wh\psi_0(\tp_1)
$$
\be     \times \prod_{i=1}^k \delta\Big(
      p_{i+1} - p_i - (\tp_{\sigma(i)+1}-\tp_{\sigma(i)})\Big)
\label{def:VA}
\ee
This complicated formula can be encoded by a  Feynman graph
and $ V_\eta(k,\sigma)$ is called the value or amplitude  of the graph.
The Feynman graph for the trivial partition 
corresponds to the usual Feynman graphs for the Gaussian case discussed in \cite{EY}
and we briefly describe their construction.
A Feynman graph consists of two directed horizontal lines (upper and lower)
with $k$ collision vertices on each that
represent the collision histories of $\bar\psi$ and $\psi$, respectively. These two
lines are  joined at the two ends. This
 corresponds to evaluating the $L^2$-norm on one end and
inserting the initial wavefunction $\psi_0$ on the other end.
Each horizontal segment carries a momentum, $p_1, p_2, \ldots p_{k+1}$
and $\tp_1, \tp_2, \ldots \tp_{k+1}$ and a corresponding (renormalized) propagator,
$(\alpha - \ov\om(p_j)-i\eta)^{-1}$ and $(\beta - \om(\tp_j)+i\eta)^{-1}$.
Here $\alpha$ and $\beta$ are the dual variables to the time on each line
and they will be integrated out, see  (\ref{def:Mi}).
Finally, the collision vertices are paired. Each pairing line
joins an upper and a lower vertex and thus can be encoded 
with a permutation $\sigma\in \cP_k$.
It is useful to think of the momenta as flowing through the lines of the graph. 
The delta function associated to each pairing line in the value of the graph (\ref{def:VA})
then expresses the Kirchhoff Law for the flow of momenta adjacent to the two vertices.

\begin{figure}
\centering
\includegraphics[height=3.5cm]{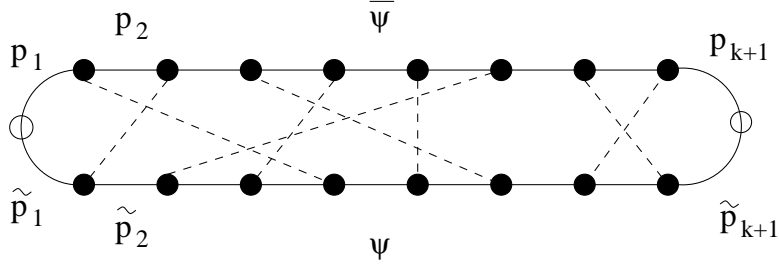}
\caption{Typical Feynman graph with no lumps}
\label{fig:1}       % Give a unique label
\end{figure}

\begin{figure}
\centering
\includegraphics[height=1.7cm]{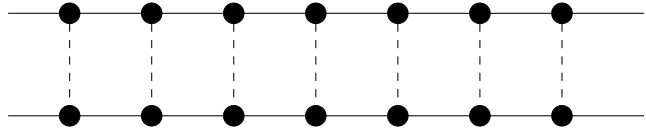}
\caption{Ladder graph}
\label{fig:2}       % Give a unique label
\end{figure}

A typical graph with trivial partition is shown  on Fig. \ref{fig:1}.
For the special case of the
 identity permutation $\sigma=id$ we obtain the so-called
 ladder graph (Fig. \ref{fig:2}).
 The following proposition shows that the
ladder gives the main contribution.

\subsubsection{The Main Contribution is the Ladder}

\begin{proposition} [$L^2$-estimate] \label{prop:L2}
Let $\eta^{-1}:=t$, $t=O(\lambda^{2+\kappa})$ and $k\leq K:=\lambda^{-\delta}(\lambda^2 t)$.
For sufficiently small $\lambda$, $\kappa$ and $\delta$  there exists a positive number
$c_1(\kappa, \delta)$ such that
\be
    \bE \|\psi_{t,k}^{(0),nr}\|^2=  V_\eta(k,id)
    + O_\delta\Big( \lambda^{c_1(\kappa, \delta)}\Big) .
\label{eq:L2bound}
\ee
%as $\lambda\ll 1$.
\end{proposition}
The threshold values for $\kappa, \delta$ and the explicit form of $c_1(\kappa,\delta)$
are found in \cite{ESY}.

\bigskip
\noindent
{\it Sketch of the proof. }
As mentioned above, we discuss only how to estimate the contributions from 
the trivial partition, but for an arbitrary
permutation $\sigma$.

\begin{figure}
\centering
\includegraphics[height=3.5cm]{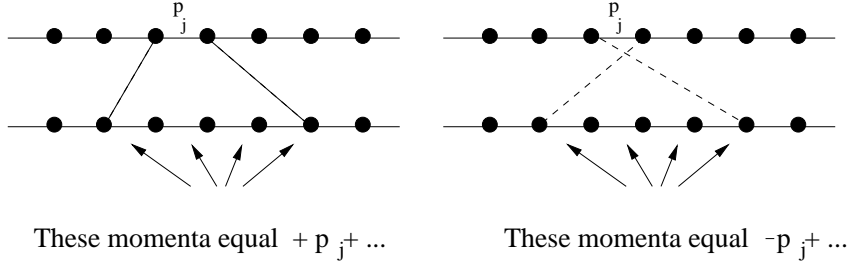}
\caption{Domain of momenta dependencies}
\label{fig:3}       % Give a unique label
\end{figure}

Given a permutation $\sigma\in \cP_k$, we define a $(k+1)\times (k+1)$
matrix $M=M(\sigma)$ as follows
\be
       M_{ij}(\sigma): = \left\{ \begin{array}{cll} 1 & \qquad \mbox{if}
    \quad & \tsi(j-1) < i \leq \tsi(j)\\
       -1 & \qquad \mbox{if} \quad & \tsi(j) <i \leq \tsi(j-1)\\
       0 & \qquad \mbox{otherwise} \quad & \;  \end{array} \right.
    \label{def:Mmat}
\ee
where, by definition, $\tsi$ is the extension of $\sigma$ to
 a permutation of $\{ 0, 1, \ldots, k+1\}$
by $\tsi(0):=0$ and $\tsi(k+1):=k+1$.
It is easy to check that
\be
    V_\eta(k,\sigma): = \frac{\lambda^{2k}e^{2t\eta}}{(2\pi)^2} \int \rd\bp  \rd\tbp\;
      M_\eta(k,\bp,\tbp)
    \prod_{i=1}^{k+1}\delta\Big( \; \tp_i - \sum_{j=1}^{k+1} M_{ij}p_j \Big)\; ,
\label{eq:deltas}
\ee
in other words, the matrix $M$ encodes the dependence of the $\tp$-momenta on
the $p$-momenta. This rule is transparent in the graphical representation
of the Feynman graph: the momentum $p_j$ appears in
those $\tp_i$'s which fall into its "domain of dependence",
i.e. the section between the image of the two endpoints of
$p_j$, and the sign depends on the ordering of these images (see Fig. \ref{fig:3})

%\bigskip\bigskip
%\centerline{\epsffile{domdep.eps}}

The matrix $M(\sigma)$ has several properties that follow easily from this structure:

\begin{lemma}\label{lemma:M}
For any permutation $\sigma\in\cP_k$ the matrix $M(\sigma)$ is

(i) invertible;

(ii) totally unimodular, i.e. any subdeterminant is 0 or $\pm 1$.
\end{lemma}

The following definition is crucial. It establishes the necessary concepts
to measure the complexity of a permutation.

\begin{definition}[Valley, peak and slope]\label{def:slopei}
Given a permutation $\sigma\in \cP_k$ let $\tsi$ be its extension. A point
 $(j, \sigma(j))$, $j\in I_k:=\{ 1, 2,\ldots, k\}$,
on the  graph of $\sigma$ is called {\bf peak} 
if $\tsi(j-1) >\sigma(j)<\tsi(j+1)$,
it is called {\bf valley} if $\tsi(j-1) <\sigma(j)>\tsi(j+1)$,
otherwise it is called {\bf slope}.
Additionally, the  point $(k+1, k+1)$
is also called valley.
%Let $I=\{ 1, 2, \ldots, k+1\}$ denote the set of row indices of $M$.
The set $I=\{ 1, 2, \ldots, k+1\}$ is partitioned
into three disjoint subsets, $I=I_v\cup I_p \cup I_s$, 
such that $i\in I_v, I_p$
or $I_s$ depending
on whether $(\tsi^{-1}(i),i)$ is a valley, peak or slope, respectively.
Finally, an index $i\in I_v\cup I_s$ is called {\bf ladder index} if
$|\tsi^{-1}(i)-\tsi^{-1}(i-1)|=1$.
The set of ladder indices is denoted by $I_\ell\subset I$
and their cardinality is denoted
by $\ell=\ell(\sigma):= |I_\ell|$.
The number of non-ladder indices, $d(\sigma): = k+1-\ell(\sigma)$ is
called the {\bf degree} of the permutation $\sigma$.
 \end{definition}

\bigskip
\noindent
{\it Remarks:} (i) The terminology of peak, valley, slope, ladder
comes from the graph of the permutation $\tsi$ viewed as a function
on $\{ 0, 1, \ldots , k+1\}$ in a coordinate system where the
vertical axis is oriented downward.

 % (see picture).

(ii) For $\sigma = id$ we have $I_p=\emptyset$, $I_s=\{1, 2,\ldots, k\}$,
$I_v=\{k+1\}$ and $I_\ell=\{1, 2, \ldots , k+1\}$. In particular, $d(id)=0$
and $d(\sigma)>0$ for any other permutation $\sigma\neq id$.

\bigskip

The following theorem shows that the degree of the permutation $d(\sigma)$
measures the size of $V_\eta(k,\sigma)$. This is the key theorem in our method
and we will sketch its proof separately in Section \ref{sec:mainsk}.

\begin{theorem}\label{thm:Vsii} Let  $\eta^{-1}:=t$, $t=O(\lambda^{2+\kappa})$
with a sufficiently small
$\kappa$. Let $\sigma\in\cP_k$
and assume that $k\leq K=\lambda^{-\delta}(\lambda^2t)$.
For sufficiently small $\kappa$ and $\delta$ there exists $c_2(\kappa, \delta)>0$ 
such that 
\be
    |V_\eta(k,\sigma)| \leq
     \Big(C\lambda^{c_2(\kappa,\delta)}\Big)^{d(\sigma)},
     \qquad \lambda\ll 1 \; .
\label{eq:Vsii}
\ee
\end{theorem}
This theorem is complemented by the following lemma:
\begin{lemma}\label{lemma:combi}
Let $k = O(\lambda^{-\kappa-\delta})$, $d>0$ integer and let
 $\gamma>\kappa+\delta$.   Then
\be
    \sum_{\sigma\in\cP_k\atop d(\sigma)\ge d} \lambda^{\gamma d(\sigma)} \leq
    O\Big(\lambda^{d(\gamma-\kappa-\delta)}\Big)
\label{eq:lambdasum}
\ee
for all sufficiently small $\lambda$.
\end{lemma}
The proof follows from the combinatorial estimate on the number of
permutations with a given degree:
$$
    \# \{ \sigma\in \cP_k\; : \; d(\sigma)=d\} \leq (Ck)^{d} \; .
$$
From  Theorem \ref{thm:Vsii} and Lemma \ref{lemma:combi}
we immediately obtain an estimate on the contribution of the
trivial lumps to $\bE \|\psi_{t,k}^{(0),nr}\|^2$ 
if $\kappa$ and $\delta$ are sufficiently small:
\be
  \sum_{\sigma\in \cP_k\atop\sigma\neq id} |V_\eta(k,\sigma)|\leq
  O_\delta\Big( \lambda^{c_3(\kappa, \delta)}\Big)
\label{nolump}
\ee
with some appropriate $c_3(\kappa, \delta)>0$.

\subsection{Sketch of the proof of the main technical theorem}
\label{sec:mainsk}

In this section we explain the proof of Theorem \ref{thm:Vsii}. We set
\be
    E_\eta(M): =
    \lambda^{2k}\!\!\iint_{-4d}^{4d} \!\!\rd\alpha \rd\beta \int\!\!\rd \bp
    \prod_{i=1}^{k+1}
    \frac{1}{|\alpha-\ov\om(p_i)-i\eta|} \prod_{j=1}^{k+1}
    \frac{1}{|\beta -\om(\sum_{\ell=1}^{k+1} M_{j\ell}p_\ell)+i\eta|}\; .
\label{def:EMi}
\ee
For the continuous model, the definition includes the $\wh B$ factors
to ensure the integrability for the large momentum regime.
It is easy to check that $V_\eta(k,\sigma)$ is
estimated by $E_\eta(M(\sigma))$ modulo constant factors and
 negligible additive terms coming from the regime where $\alpha$ or $\beta$ is big.

The denominators in this multiple integral are almost singular
in certain regimes of the high dimensional space of all momenta.
The main contribution comes from the overlap of these singularities.
The overlap structure is encoded in the matrix $M$, hence in
the permutation $\sigma$, in a very complicated entangled way.
Each variable $p_j$ may appear in many denominators in (\ref{def:EMi}),
so successive integration seems very difficult. We could not find
the exact order (as a power of $\lambda$)
of this multiple integral but we conjecture that true order is 
essentially $\lambda^{2 d(\sigma)}$.
Our goal in Theorem \ref{thm:Vsii} is to give a weaker 
bound of order $\lambda^{cd(\sigma)}$, i.e. that
is still a $\lambda$-power linear in the degree, but the coefficient
considerably smaller than 2.

Notice that the $\alpha$-denominators in (\ref{def:EMi}) correspond to the
columns of $M$ and the $\beta$-denominators corresponds to the rows.
For this presentation we will use $j$ to label row indices and $i$
to label column indices.
We recall the sets  $I_v, I_p, I_\ell$ from Definition \ref{def:slopei}
and we will view these sets as subsets of the row indices of $M$.

First we notice that if $j\in (I_\ell\setminus I_v)$, i.e. $j$ is
a non-valley ladder row, then there exists
a column index $i=c(j)$ such that the momentum $p_i$
appears only in the $j$-th $\beta$-denominator. In
other words, the $i$-th column of $M$ has a single
nonzero element (that is actually $\pm 1$)  and it is
in the $j$-th row. Therefore the $\rd p_i$ integral
can be performed independently of the rest of the integrand by
using the following elementary but quite involved  bound for small $\kappa$:
\be
  \sup_{w, \alpha,\beta} \int_{\Tor^d} \rd p_i
 \frac{\lambda^2}{|\alpha - \ov\om(p_i)-i\eta|\;
  |\beta - \om(\pm p_i + w)+i\eta|} \leq 1 + O(\lambda^{1/4})\; .
\label{eq:ladd}
\ee
Note that the constant of the main term is exactly 1. This fact is important,
since in graphs with low degree this estimate has to be raised
to a power $|I_\ell\setminus I_v|$ that may be comparable with $k$. Clearly
for $k\leq K\sim \lambda^{-\kappa-\delta}$ and $\kappa+\delta<1/4$ we have
\be
     \Big( 1 + O(\lambda^{1/4}) \Big)^k \leq const \; ,
\label{cc}
\ee
but had 1 been replaced with a bigger constant in (\ref{eq:ladd}),
we would obtain an exponentially big factor  $(const)^k$ that would  not  be affordable.
The precise constant 1 in the estimate (\ref{eq:ladd})
 is related to the appropriate choice of the renormalization $\theta(p)$
in $\om(p)$.

\bigskip

After the non-valley ladder rows are integrated out, and the corresponding
rows and columns are removed from the matrix $M$, we obtain a smaller
matrix $M^{(1)}$ describing the remaining denominators.
 In $M^{(1)}$ we keep the original
labelling of the rows from $M$.

Now we estimate some of the  $\beta$-denominators
in (\ref{def:EMi}) by $L^\infty$ norm, i.e. by $\eta^{-1}$.
This is a major overestimate, but these denominators
are chosen in such a way that the entangled structure
imposed by $M$  becomes much simpler and many other
denominators can be integrated out by $L^1$-bounds that are
only logarithmic in $\lambda$.

We start with estimating all $\beta$-denominators in rows $j\in I_p$
by the trivial $L^\infty$-norm. 
The corresponding rows are removed from $M^{(1)}$,
in this way we obtain a matrix $M^{(2)}$.
Let
$$
I^*:= I\setminus \Big( I_p \cup (I_\ell\setminus I_v) \Big)
$$
be the remaining row indices
after removing the peaks and the non-valley ladders.

Then we inspect the remaining
rows $j\in I^*$ of $M^{(2)}$
in increasing order.
 The key observation is that
for each $j\in I^*$  there exists
a column index, $i=c(j)$, such that the variable $p_i$
appears {\bf only}  in the $j$-th $\beta$-denominator, provided
that all $\beta$-denominators with $j'<j$ have already
been integrated out. In view of the structure of $M^{(1)}$,
it means that for any $j\in I^*$
there exists a column $i=c(j)$ such that the only
nonzero element among $\{ M_{ij'}^{(2)}\; : \; j'\ge j\}$
is $M_{ij}^{(2)}$. This fact follows from the structure of $M(\sigma)$
and from the fact that all rows with $j\in I_p$ have been removed.

This property allows us
to remove each remaining $\beta$-denominator, one by one,
by estimating integrals of the type
\be
   \int_{\Tor^d} \rd p_i \; \frac{1}{|\alpha - \ov\om(p_i)-i\eta|} \;
   \frac{1}{|\beta - \om(\pm p_i + w)+i\eta|} 
\leq \frac{C \eta^{-\tau}}{|w|}  \; ,
\label{eq:2den}
\ee
where $w$ is a linear combination of momenta other than $p_i$.
The absolute value $|w|$ is interpreted as the distance of $w$
from the nearest critical point of the dispersion relation $e(p)$.
The variable $p_i$ at this stage of the procedure appears
only in these two denominators. 

The exponent $\tau$ can be chosen zero (with logarithmic corrections)
for the continuous model and this fact has already been used in \cite{EY}.
For the discrete model we can prove (\ref{eq:2den}) with $\tau=3/4+2\kappa$
and we know that the exponent cannot be better than 1/2. 
The reason for the weaker estimate is the lack of convexity
of the level set $\Sigma_e$. Replacing $\om(p)$ with $e(p)$
for a moment, the inequality (\ref{eq:2den}) with $\tau=0$ essentially states that
the level set
$\{ \alpha = e(p)\}$ and its shifted version $\{ \beta = e(p+w)\}$
intersect each other transversally, unless $w$ is close to zero.
Indeed, the transversal intersection guarantees that the volume of
the $p$ values, where {\it both} denominators are of order $\eta$,
is of order $\eta^2$. Then a standard argument with dyadic
decomposition gives the result with a logarithmic factor.
For translates of spheres the transversal intersection property
holds, unless $w\sim 0$. However, in certain points of the level sets $\Sigma_e$ of
the discrete dispersion relation the curvature vanishes, in fact
$\Sigma_e$ even contains straight lines for $2\leq e \leq 4$. The
transversal intersection fails in certain regions and results in a weaker bound.

Neglecting the point singularity $|w|^{-1}$  in (\ref{eq:2den}) for a moment
(see Section \ref{sec:point} later),
we easily see that with this algorithm one can bound
$E_\eta(M(\sigma))$ by
$\lambda^{2(k-q)} \eta^{-p}\eta^{-\tau(k-p-q)}$,
 modulo logarithmic factors, where
$p=|I_p|$ is the number of peak indices and $q:=|I_\ell\setminus I_v|$
is the number of non-valley ladder indices.
From the definitions it follows
that the sets $I_v$, $I_p$ and $I_\ell\setminus I_v$ are
 disjoint  and $|I_v| = p+1$.
Thus  we have  $2p+1 + q \leq k+1$. Therefore
\be
\lambda^{2(k-q)} \eta^{-p(1-\tau) -\tau(k-q)} \leq 
(\lambda^4 t^{\tau+1})^{(k-q)/2} \leq
 (\lambda^4 t^{\tau+1})^{\frac{1}{2}
[d(\sigma)-1]}
\label{lt}
\ee
since $q\leq \ell$. If $\tau<1$, then with a sufficiently small $\kappa$
we see that $\lambda^4 t^{\tau+1}$ is a positive power of $\lambda$. Thus we obtain
a bound where the exponent of $\lambda$ is linear 
in $d(\sigma)$.  With a more careful estimate one can remove
the additional $-1$ in the exponent. In particular, for the continuous
case with $\tau=0$ this argument works up to $\kappa<2$.

\bigskip

We end this section with a remark. Apparently
 the bound $\kappa <2$ (or, equivalently, $t\ll \lambda^{-4}$)
shows up in two different  contexts in this argument.
To avoid misunderstandings, we explain briefly that neither of these two
appearences is the genuine signature of the expected threshold $\kappa=2$
for our expansion method to work.
The true reason is the one mentioned in the introduction:
even the best possible bound, $\lambda^{2d(\sigma)}$,
on the graph with permutation $\sigma$, cannot beat the $k!$
combinatorics of the graphs beyond $\kappa =2$.

In the argument above,
on one hand, $\kappa<2$  is related to the error term in the ladder calculation
(\ref{eq:ladd}). This error term can be improved to
$\lambda^2|\log\lambda|$ and it is apparently due to the fact
that the renomalization term $\theta(p)$ was solved only
up to lowest order. An improvement may be possible
by including more than the lowest order of the self--energy.

The second apperance of $\kappa<2$, or $t\ll \lambda^{-4}$, at least 
for the continuous model, is
 in (\ref{lt}) and it  is due to the fact
that certain $\beta$-denominators are overestimated by $L^\infty$.
This is again a weakness of our method; we did overestimates
in order to simplify the integrand.

\subsection{Point singularities}\label{sec:point}

The argument in the previous section has neglected the
point singularity arising from (\ref{eq:2den}). While a point
singularity is integrable in $d\ge 3$ dimensions, it may happen that
exactly the same linear combinations of the independent variables
keep on accumulating by the repeated use of the bound (\ref{eq:2den}).
In that case at some point a high negative power of $|w|$
needs to be integrated. While it is possible to improve  the estimate
 (\ref{eq:2den}) by changing 
the denominator
on the right hand side  to $|w|+\eta$, this would
still yield further negative $\eta$-powers.

It is easy to see that this phenomenon does occur. Primarily this would have
occurred if we had not treated the ladders separately: if $p_i$'s are
ladder variables, then the corresponding $w$ momenta in (\ref{eq:2den})
are indeed the same. Although we have removed the ladders beforehand,
the same phenomenon occurs in case of a graph which contains ladder {\it only
as a minor but not as a subgraph}. Our separate ladder integration
procedure (\ref{eq:ladd}) can be viewed as a very simple renormalization
of the ladder subgraphs. The correct procedure should renormalize all
ladder minors as well.

To cope with this difficulty, we have to follow more precisely the
point singularities. To this end, we define the following
generalization of $E_\eta(M)$.
For any index set $I'\subset I=\{ 1, 2,\ldots, k+1\}$, any $|I'|\times(k+1)$
matrix $M$, any $\nu$ integer
 and any $\nu\times(k+1)$ matrix $\cE$ we define
$$
    E_\eta(I', M, \cE): = \lambda^{2k}e^{2t\eta}
    \iint_{-4d}^{4d} \rd\alpha \rd\beta
    \int \rd \bp
$$
\be
\times\Bigg(\prod_{i\in I'}
    \frac{1}{|\alpha-\ov\om(p_i)-i\eta|}\frac{1}{|\beta -\om\big(
    \sum_{j=1}^{k+1} M_{ij}p_j \big)+i\eta|}\Bigg)
    \prod_{\mu=1}^{\nu} \frac{1}{|\sum_{j=1}^{k+1}\cE_{\mu j}p_j |}
\label{eq:EE}
\ee
We follow the same procedure as described in Section \ref{sec:mainsk},
but we also keep track of the evolution of the point singularity matrix $\cE$.
At the beginning $I'= I$, $\nu=0$ and $\cE$ is not present. After the first
non-ladder type integration, a point singularity will appear from
(\ref{eq:2den}). Some of the  point singularities may get integrated
out later as one of their variables become integration variable.
Therefore we will need the following generalization of  (\ref{eq:2den}):

\begin{lemma} There exists a constant $C$ such that for any index set $A$
$$
    \sup_{|\alpha|, |\beta|\leq 4d} \int
    \frac{1}{|\alpha - \ov\om(p)-i\eta|\; |\beta - \om(r+p)+i\eta|}
    \prod_{a\in A} \frac{1}{|r_a+p|} \;   \rd p \;
$$
\be
    \leq C\eta^{-\tau'} |\log\eta|^3 \sum_{a\in A} \Big(\prod_{a'\in A\atop a'\neq a}
    \frac{1}{|r_a-r_{a'}|} \Big) \frac{1}{|r|}\; . \qquad \qed
\label{eq:dpi}
\ee
For the continuous model $\tau'=0$ while for the discrete model $\tau'= \frac{7}{8}+2\kappa$.
\end{lemma}
Using this lemma, we can keep record of the evolution of the
point singularity matrix $\cE$ at an intermediate step
of our integration algorithm. These matrices change by simple operations
reminiscent to the
Gaussian elimination.

\bigskip

Three  complications occur  along this procedure, we briefly describe
how we resolve them:

\medskip

(1) The inequality (\ref{eq:dpi})
 does not allow higher order point singularities.
Although it is possible to generalize it to include such singularities as well,
we followed a technically simpler path. In addition to the indices $j\in I_p$,
we select  further $\beta$-denominators that we estimate by the
trivial $L^\infty$ bound. These additional
indices are chosen in such a way, that (i) the number of remaining
rows be at least $\frac{1}{3} d(\sigma)$; (ii) the point singularity
matrix be of full rank at every step of the algorithm. This second
criterion  guarantees that no higher order point singularities
occur. Since every $\cE$ can be derived from $M$
by a  procedure that is close to Gaussian elimination and $M$ is
invertible (Lemma \ref{lemma:M}),
the full-rank property is relatively easy to guarantee.

\medskip

(2) The full-rank property actually needs to be guaranteed in
a quantitative way, at least  the entries of $\cE$ needs to be
controlled. These entries  appear in the point singularity
denominators of (\ref{eq:dpi}) and their inverses would appear
in the estimate. The key observation is that each entry of every matrix
$\cE$ along the procedure is always 0, 1 or $-1$. It is actually easier to prove
a stronger statement, namely that every $\cE$ is a {\it totally unimodular matrix}.
The proof follows from
the fact that every $\cE$ can be derived from $M$ by elementary Gaussian elimination
steps plus zeroing out certain rows and columns. Such steps preserve
total unimodularity and $M$ is totally unimodular by Lemma \ref{lemma:M}.

\medskip

(3) After all $\beta$-denominators are eliminated, we are
left with an integral of the form
\be
     E_\eta(J, \emptyset, \cE): = \int_{-4d}^{4d}
    \rd \alpha \int \Big(\prod_{i\in J}\rd p_i\Big) \prod_{i\in J}
    \frac{1}{|\alpha - \ov{\om }(p_i) -i\eta|}
    \prod_{\mu=1}^\nu \frac{1}{|\sum_{i\in J}\cE_{\mu i}p_i|}\; 
\label{eempty}
\ee
for some index set $J$ and some point singularity matrix obtained
along the integration procedure.
Without the point singularities, this integral could be estimated
by the $|J|$-th power of $|\log \eta|$. Since $\cE$ is
totally unimodular, a similar estimate can be obtained for (\ref{eempty})
as well.

\section{ Computation of the main term and its convergence to a Brownian
motion}

Our goal is to compute the Wigner distribution
$\bE  W^\e_{\psi_{t}} (X, v)$ with $t=\lambda^{-2-\kappa}T$ and $\e =\lambda^{2+\kappa/2}$.
From Proposition \ref{prop:L2}, and similar bounds on
the repetitive terms in (\ref{eq:duha}),
we can  restrict our attention
to the ladder graph. 
The following lemma is a more precise version of the ladder integration
Lemma (\ref{eq:ladd}) and it is  crucial
to this computation. We present it for the more complicated
discrete case. The proof is a tedious 
calculation in \cite{ESY}.

\begin{lemma}\label{lemma:opt1}
Suppose $f(p)$ is
a $C^1$ function on $\Tor^d$.
Recall $0<\kappa < 1/16$ and define $\gamma: = (\a +\beta)/2$.
Let  $\eta$ satisfy
 $\lambda^{2+ 4\kappa}\leq \eta\leq \lambda^{2+\kappa}$.
Then for $|r|\le \lambda^{2+\kappa/4}$ we have,
\be
 \int \frac {\lambda^2 f(p) }{( \a - \ov\om(p-r)
       - i\eta )
     (\beta - \om(p+r)  +i\eta )} \; \rd p   \qquad\qquad\qquad
\ee
$$
     =   -2\pi i\lambda^2\int   \frac{ f(p)\; \delta(e(p)-\gamma)}{ (\a-\beta)
 + 2 (\nabla  e)(p) \cdot r -
       2 i [\lambda^2  {\rm Im}\Theta (\gamma)
       +\eta ]} \, \rd p + O(\lambda^{1/2-8\kappa}|\log\lambda|) \; .
$$
\end{lemma}

 Since the Boltzmann collision kernel is uniform
on the energy shell, the calculation of $\bE  W^\e_{\psi_{t}} (X, v)$
is more straightforward for the
discrete case. We present the sketch of this calculation,
the continuous model requires a little more effort at this stage.

Let $\e =\lambda^{2+\kappa/2}$ be the space scale.
 After rescaling the Wigner function at time $t$, we compute
$\wh W(\e\xi, v)$ tested against a smooth, decaying function
$\cO(\xi, v)$. In particular  $\xi$ is of order 1.
After the application of Lemma \ref{lemma:opt1} (with $v=v_{k+1}$)
and change of variables  $a:=(\a+\beta)/2$ and $b:=\lambda^{-2}(\a-\beta)$, we obtain
$$
\langle \cO,\bE \wh W\rangle:= \int \rd v\rd \xi \; \cO(\xi, v)\bE \wh W(\e\xi, v) 
  =\sum_{k\leq K} 
\iint_\bR \frac{\rd \alpha \rd \beta}{(2\pi)^2\lambda^2} \int \rd\bv \; 
e^{it(\alpha-\beta)+2\eta t} 
$$
$$
\times \cO(\xi, v_{k+1})\wh W_0(\e\xi, v_1)\prod_{j=1}^{k+1} 
     \frac{\lambda^2}{\Big(\alpha - \ov\om(v_i +\frac{\e\xi}{2}) - i\eta\Big)
   \Big(\beta - \om(v_i -\frac{\e\xi}{2})  + i\eta\Big)} 
$$
$$
  \approx \sum_{k\leq K} \iint_\bR \frac{\rd a \rd b}{(2\pi)^2}  \; 
e^{it\lambda^2b}\!\! 
     \int \Big(\prod_j  
    \frac{-2\pi i \; \delta(e(v_j)-a)\rd v_j }{ b + \lambda^{-2}\e\nabla e(v_j)\cdot \xi -
       2 i \cI (a)} \Big) \wh W_0(\e\xi, v_1)\cO(\xi, v_{k+1})\; ,
$$ 
where we defined $\cI(\gamma):= {\rm Im}\Theta (\gamma)$
for brevity.
We used $\eta=\lambda^{2+\kappa}$ to estimate the error terms.
The main term (left hand side above)  however, is independent of
$\eta$, so we can choose $\eta=\lambda^{2+4\kappa}$ for the 
rest of the calculation and we note that  Lemma \ref{lemma:opt1}
holds for this smaller $\eta$ as well. This is the reason why
the $e^{2\eta t}$ factor is negligible.

We expand the fraction up to second order in $\e$, we get
$$
   \frac{-i }{ b +  \lambda^{-2}\e\nabla e(v_j)\cdot \xi -
       2 i  \cI(a) } \approx \frac{- i}{ b  -
       2 i          \cI(a) }
   \Bigg[ 1 - \frac{  \lambda^{-2}\e\nabla e(v_j)\cdot \xi}{b- 2 i  \cI(a)}
     +  \frac{ \lambda^{-4}\e^2 [\nabla e(v_j)\cdot \xi]^2}{ (b- 2 i
        \cI(a) )^2}\Bigg] \nonumber
$$
By symmetry of the measure $2\pi\delta(e(v)-a)\rd v$ under the sign flip, $v\to-v$ and
using  $(\nabla e)(v)=-\nabla e(-v)$, we see that the first 
order term vanishes after the integration.
We also define the matrix
$$
    D(a):= \frac{1}{2\,  \cI(a)} \int \rd \mu_a(v)\; \frac{\nabla e(v)}{2\pi}
   \otimes \frac{\nabla e(v)}{2\pi}
$$
After  integrating out all momenta and changing the  $b$ variable we obtain
$$
\langle \cO, \bE \wh W\rangle
\approx  \sum_{k\leq K} \int \rd\xi \int_\bR  \frac{2\cI(a)\rd a}{2\pi}
\langle \cO(\xi, \cdot )\rangle_a \langle\wh W_0(\e\xi, \cdot )\rangle_a
\int_\bR \frac{\rd b}{2\pi}  \; e^{2i\lambda^2 tb
 \cI(a)}
$$
$$
   \qquad\qquad\qquad\times  \Big(\frac{-i}{ b  -i}\Big)^{k+1} 
 \times \Bigg[ 1 
     + \frac{ (2\pi)^2\e^2\lambda^{-4}
       \langle \xi, D(a)\xi\rangle}{2\cI(a)}\cdot\frac{1}{(b- i)^2}\Bigg]^{k+1}
$$
We  sum up the geometric series and perform a residue calculation to evaluate
the $\rd b$ integral. We obtain that
the main contribution comes from $k\sim  2\lambda^2 t  \cI(e) $,
so the truncation $k\leq K$ can be neglected and  the result 
of the $\rd b$ integration is 
$$
 \int_\bR \frac{\rd b}{2\pi}  \Big(\cdots\Big)
  \approx \exp \Bigg( -  \frac{ (2\pi)^2\e^2 t \langle \xi, 
 D(e)\xi\rangle}{ \lambda^2}\Bigg)
$$
To obtain a nontrivial limit, $\e^2t/\lambda^2\sim 1$ is necessary.
Noting that $t=\lambda^{-2-\kappa}T$ with $T=O(1)$,
we see that indeed the space must be scaled by $\e = \lambda^{2+\kappa/2}$.
Finally we obtain
$$
  \langle \cO, \bE \wh W\rangle
  \approx  \int \rd\xi  \int_\bR  \frac{2\cI(a)\rd a}{2\pi}
\langle \cO(\xi, \cdot )\rangle_a \langle\wh W_0(\e\xi, \cdot )\rangle_a
 \exp\Big( -  (2\pi)^2 T\langle \xi, D(a)\xi\rangle \Big)
$$
Since $ \exp[ -  (2\pi)^2 T\langle \xi, D(a)\xi\rangle ]$ is
the fundamental solution to the heat equation (\ref{eq:heat}),
from the definition of $\langle \cdot \rangle$, and 
after inverse Fourier transform
we obtain  (\ref{fint}).
This completes the sketch of the calculation of the main term.

\bigskip

{\bf Acknowledgements.} L.Erd\H os was partially supported by NSF grant 
DMS-0200235 and EU-IHP Network ``Analysis
and Quantum'' HPRN-CT-2002-0027. H.-T. Yau was  partially supported
by NSF grant DMS-0307295 and MacArthur Fellowship.

\printindex
\end{document}